\documentclass{aastex}
\usepackage{natbib}
\usepackage{emulateapj5}

\slugcomment{Submitted to Astrophysical Journal Letters}

\shorttitle{PSR J2021+3651}
\shortauthors{Roberts et al.}

\begin{document}
\title{PSR J2021+3651: A YOUNG RADIO PULSAR COINCIDENT WITH AN UNIDENTIFIED EGRET $\gamma$-RAY SOURCE}

\author{Mallory S. E. Roberts\altaffilmark{1},
Jason W. T. Hessels,
Scott M. Ransom\altaffilmark{1},
Victoria M. Kaspi\altaffilmark{1}}
\affil{Physics Department, McGill University, Rutherford Physics Building,
3600 University Street, Montreal, QC, H3A 2T8, Canada}
\email{roberts@physics.mcgill.ca}

\author{Paulo C. C. Freire}
\affil{NAIC, Arecibo Observatory, HC03 Box 53995, PR 00612, USA}
\author{Fronefield Crawford}
\affil{Physics Department, Haverford College, Haverford, PA 19041, USA}
\and
\author{Duncan R. Lorimer}
\affil{University of Manchester, Jodrell Bank Observatory, Macclesfield, Cheshire SK11 9DL, UK}
\altaffiltext{1}{Center for Space Research, Massachusetts Institute of Technology, Cambridge, MA 02139, USA}

\begin{abstract}
We report on a deep search for radio pulsations 
toward five unidentified {\it ASCA} X-ray sources coincident 
with {\it EGRET} $\gamma$-ray sources.
This search has led to the discovery of a young and energetic pulsar 
using data obtained with the new Wideband Arecibo Pulsar Processor. 
PSR~J2021+3651 is likely associated 
with the X-ray source AX J2021.1+3651, which in turn is likely associated with 
the {\it COS~B} high energy
$\gamma$-ray source 2CG~075+00, also known as GeV~J2020+3658 or 3EG~J2021+3716. PSR~J2021+3651 
has a rotation
period $P\cong 104$~ms and $\dot P \cong 9.6\times10^{-14}$, implying 
a characteristic age $\tau_c\sim  17$ kyr and a spin-down luminosity 
$\dot E\sim3.4\times 10^{36}$ ergs~s$^{-1}$. The dispersion measure DM$\simeq
371$~pc cm$^{-3}$
is by far the highest of any observed pulsar in the Galactic longitude range 
$55^{\circ} < l < 80^{\circ}$.  This DM suggests a distance $d\ga 10$~kpc, 
and a high $\gamma$-ray efficiency of $\sim$ 15\%, but the true distance may 
be closer if there is a significant contribution to the DM from
excess gas in the Cygnus region. 
The implied luminosity of the 
associated X-ray source suggests the X-ray emission is dominated by a 
pulsar wind nebula unresolved by {\it ASCA}. 
\end{abstract}


\section{Introduction}

The majority of high energy $\gamma$-ray sources observed by {\it EGRET} 
and other telescopes have long escaped identification with lower energy
counterparts \citep{hbb+99}.  Young pulsars remain the only Galactic source class (other than the Sun) unambiguously
shown to emit radiation in the 100~MeV -- 10~GeV range \citep{tho01}. It is likely that many of
the unidentified $\gamma$-ray sources at low Galactic latitudes are young
pulsars as well. Many of these sources have characteristics similar to 
those of the known $\gamma$-ray pulsars, but have no known pulsars within their error boxes.
This fact, along with modelling of the multi-wavelength pulse profiles
and the still singular example of Geminga \citep{hh92,hel94}, has led to the suggestion that
a large fraction of the radio beams from $\gamma$-ray sources will miss
the Earth and appear radio quiet \citep{rom96a}.

Recently, a number of young pulsars coincident with known $\gamma$-ray
sources have been discovered \citep{dkm+01_mal,cbm+01}. These new discoveries are largely a 
result of greater sensitivity to pulsars with high dispersion measures (DM) 
obtainable
with newer pulsar backends such as the Parkes multibeam system \citep{mlc+01}. 
The recent detection of a young radio pulsar
in the supernova remnant 3C58 with a 1400~MHz flux density of only
$\sim 50$~$\mu$Jy \citep{csl+02} suggests many more faint radio pulsars 
await discovery in deep, targeted searches.

A major stumbling block in the identification of the {\it EGRET} sources is their 
large positional uncertainty, which can be greater than $1^{\circ}$
across.  We approach this problem by targeting
potential hard X-ray counterparts, whose size and positional
uncertainty are much smaller than the typical single dish radio beam. 
Using as our guide the {\it ASCA} catalog of potential X-ray counterparts of 
GeV sources (based on the \citet{lm97} catalog of
sources with significant flux above 1 GeV) by \citet{rrk01} 
(hereafter, RRK), we have
searched five X-ray sources for radio pulsations using the 305-m Arecibo telescope
and the 64-m Parkes telescope (see Table~\ref{tab:observations}).
Previous searches of these targets were limited.  In particular, two of the 
three sources
observed at Parkes (AX J1418.7$-$6058 and AX J1809.8$-$2332)
were not previously the subject of any directed search and were
only observed as a matter of course during the Parkes Multibeam Galactic Plane Survey \citep{mlc+01}.
A survey of {\it EGRET} sources by \citet{ns97} looked at two of the sources
searched here (AX J1826.1$-$1300 and AX J2021.1+3651) with a limiting flux density for 
slow pulsars of 0.5 -- 1.0~mJy at frequencies of 370 and 1390~MHz, but found no new pulsars.
Our search has led to the discovery of one young and energetic pulsar, PSR J2021+3658.
We argue that it is a likely counterpart to AX J2021.1+3651 and GeV J2020+3658 / 2CG 075+00.

\section{Observations and Analysis} 

\subsection{Observations}

On 2002 January 30 and 31, we observed the only two unidentified 
sources in the RRK catalog visible from the Arecibo radio 
telescope, AX J1907.4+0549 and AX J2021.1+3651, 
using the Wideband Arecibo Pulsar Processor (WAPP).  
The WAPP is a fast-dump digital
correlator with adjustable bandwidth (50 or 100~MHz) and variable numbers 
of lags and sample times (for details see Dowd, Sisk, \& Hagen 
2000\nocite{dsh00}).
Our observations were made at 1.4~GHz with 100~MHz of bandwidth and summed 
polarizations.  The observational parameters are summarized in 
Table~\ref{tab:observations}.  The 16-bit samples were written to a disk array 
and then tranfered to magnetic tape for later analysis.

On 2001 February 11$-$15, the three extended hard X-ray sources listed by RRK
as potential pulsar wind nebulae,  AX J1418.7$-$6058 (the Rabbit),
AX J1809.8$-$2333, and
AX J1826.1$-$1300, were searched for radio pulsations with the Multibeam 
receiver on the Parkes radio telescope.  Each source was observed once at a 
central observing frequency of 1390~MHz with a
512 channel filterbank spectrometer covering 256~MHz of bandwith, and once at a central observing frequency of 1373~MHz with 96 channels and 288~MHz of 
bandwith (see Table~\ref{tab:observations}).
During each observation, signals from each channel were square-law 
detected and added in polarization pairs before undergoing high-pass filtering.
The signals were one-bit digitized every 0.25~ms and recorded onto magnetic tape for 
later analysis. 
                                                                                                  
\subsection{Analysis}

Analysis of Arecibo observations was done using the {\tt PRESTO} software suite \citep{ran_thesis}
by first  
removing obvious narrow band and/or short duration interference in both
the time and frequency domains.  We
then dedispersed the data at 500 trial DMs between 10 and 510~pc cm$^{-3}$ 
for AX J2021.1+3651 and 540 trial DMs between 0 and 2695~pc cm$^{-3}$ for AX J1907.4+0549. 
Employing harmonic summing, the FFTs of each time series were searched, and 
interesting candidates were folded over a fine grid 
in DM, period, and period derivative space to optimize the signal-to-noise.

The Parkes observations  were analyzed using standard pulse search software 
and a similar procedure by searching 
the 96 channel data at 279 trial DMs ranging from 0 to 1477~pc
cm$^{-3}$ and the 512 channel data at 501 trial DMs ranging from 0 to 670~pc
cm$^{-3}$. 
We tested the system by observing a known bright pulsar 
(PSR B1124$-$60) for 300 s, which was clearly
detected in the processing. Re-analysis of the data using {\tt PRESTO}
has not revealed any new candidates.

\section{RESULTS}

\subsection{PSR J2021+3651}

\begin{figure}
\plotone{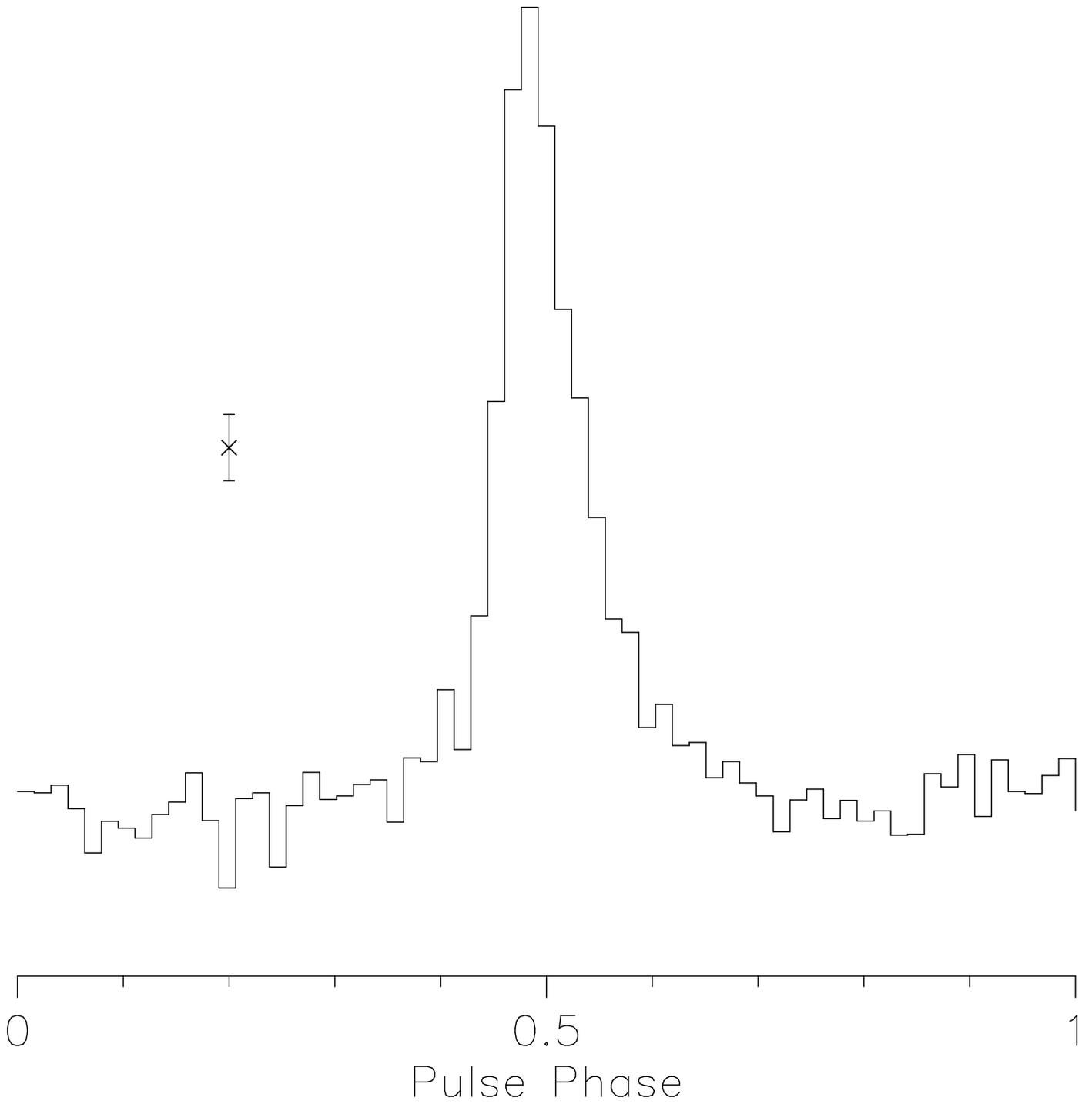}
\figcaption{1.4~GHz pulse profile for PSR J2021+3651 from the MJD 52305 observation. The error bar represents the 1~$\sigma$ uncertainty.  \label{fig:pulse}}
\end{figure}

A new highly dispersed 104-ms pulsar 
was detected in the Arecibo observations made of AX J2021.1+3651;
it is clearly visible in both of the original search observations, and represents a $\sim$ 40 $\sigma$
detection in the longest observation.  The pulse
profile is shown in Figure~\ref{fig:pulse}.  

A subsequent series of 7 observations performed between MJD 52405$-$52416 
allowed us to determine a phase connected solution for some of the
pulsar parameters.
Integrated pulse profiles from these observations were convolved with a template
profile to extract 12 topocentric times of arrival (TOA).  Using 
{\tt TEMPO}\footnote{See http://pulsar.princeton.edu/tempo} and adopting the
ROSAT position for the pulsar (\S 4.1), the topocentric TOAs 
were converted to TOAs at the solar system barycenter at infinite frequency 
and fit simultaneously for pulsar period, period derivative, 
and DM, with a residual rms of 91~$\mu$s.  The measured and derived parameters
for this pulsar are listed in Table~\ref{tab:pulsar}.

\subsection{Non-detections}

No convincing pulsar candidates were detected in any of the search
observations conducted at Parkes. We estimate upper limits of 
$S \cong 0.08$ mJy at 1.4~GHz for pulse                     
periods $P$ $\ga$ 10 ms for most of the 512 channel observations. A comparable 
sensitivity was obtained for long periods in the 96 channel
observations. For DMs larger than about 100~pc cm$^{-3}$, the sensitivity 
to fast pulsars (P $\la$ 50 ms) is significantly degraded in the 96 
channel system. These sensitivity limits were estimated using a
sensitivity modeling technique described in detail elsewhere (e.g., 
Manchester et al. 2001\nocite{mlc+01}).

Likewise, extensive searching of AX J1907.4+0549 yielded no convincing pulsar candidates.   
We estimate an upper limit of $S\simeq 0.02$~mJy 
at 1.4~GHz for a long period pulsar assuming a 10\% duty cycle.

\section{Discussion}

\subsection{PSR J2021+3651}

Our search targeted AX J2021.1+3651, which was identified as a potential high-energy counterpart to GeV J2020+3658 by RRK.  
The X-ray source is near the
{\it ASCA} field edge and so the positional uncertainty from {\it ASCA} is $\ga 
1^{\prime}$ \citep{guf+00}. 
A subsequent search of the {\it ROSAT} All-Sky Survey Faint Source
catalog \citep{vab+00} revealed the source 1RXS J202104.5+365127 with
a smaller positional error of $24^{\prime \prime}$.  
Given the rarity of such young, energetic pulsars and the small
size of the Arecibo beam (3$^\prime$ at FWHM),
an association with the X-ray source is highly probable.

The DM of PSR J2021+3651 is by far the highest known in the 
Galactic longitude 
range $55^{\circ} < l < 80^{\circ}$ which is mainly an inter-spiral arm 
direction. The \citet{tc93} model gives a distance
of $\sim 19$ kpc, well beyond the last spiral arm used in the 
model.  A revised model is currently in preparation by Cordes and Lazio which
includes an outer spiral arm at $d\sim 10$~kpc.  Placing the pulsar at the far 
edge of 
this outer arm still does not account for all the
observed dispersion; and it is possible that there are further contributions
from clouds in the Cygnus region, 
where there is known to be excess gas at $d\sim 1.5$~kpc (J. Cordes,
private communication).  However, there are no obvious HII regions 
within the Arecibo beam seen in either Very Large Array (VLA) 20-cm radio or
Midcourse Space Experiment (MSX) 8.3~$\mu$m images (available from 
the NASA/IPAC Infrared Science Archive).  

\begin{figure}
\plotone{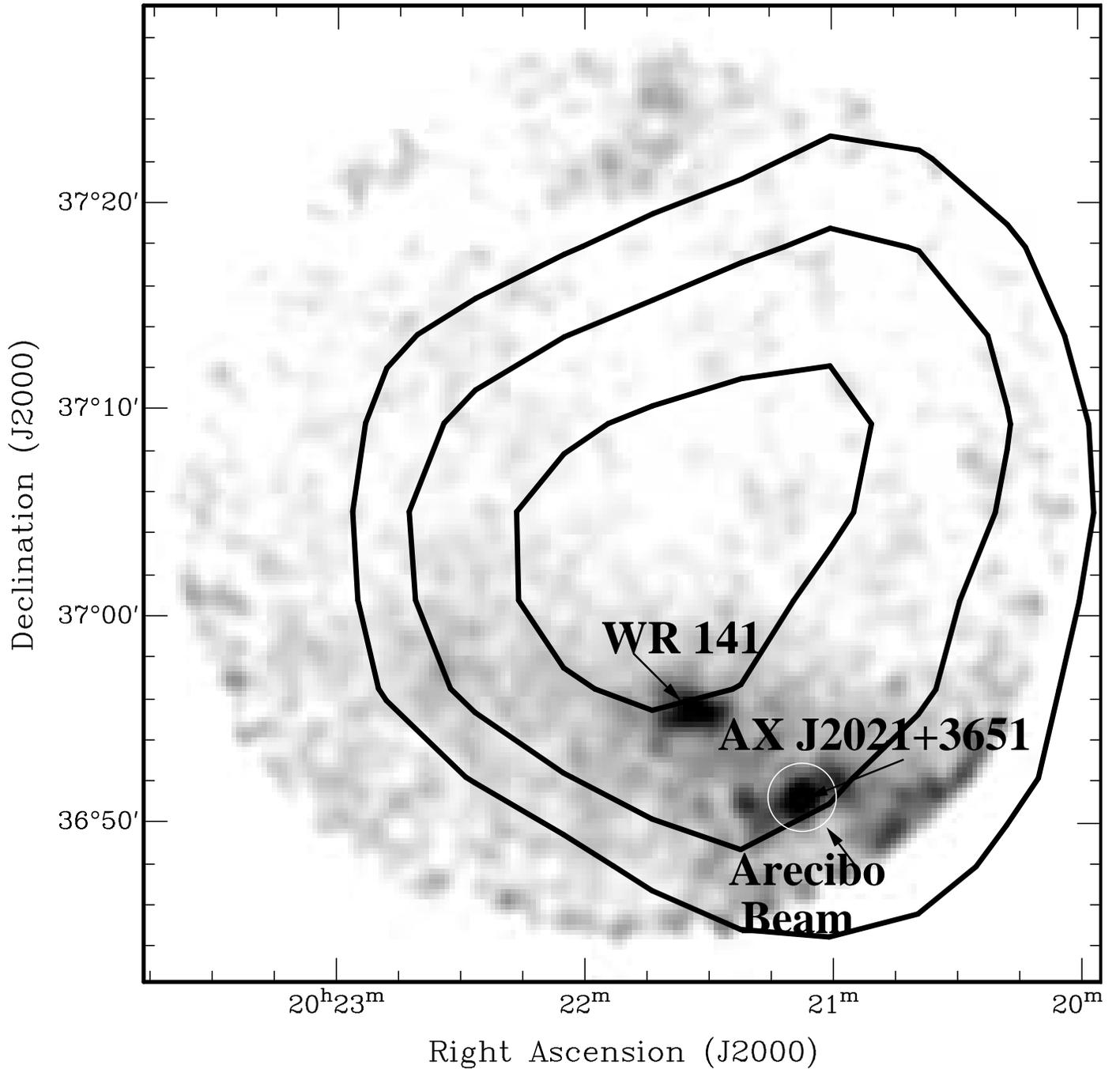}
\figcaption{ASCA GIS 2-10 keV image of the $\gamma$-ray source region.  
The contours are the 68\%, 95\%, and 99\% confidence
regions of the $\gamma$-ray source position, derived from the $>$ 1~GeV photons
\citep{rrk01}.  WR 141 is a Wolf-Rayet star also in the field of view.  The circle
centered on AX J2021.1+3651 indicates the size of the $3'$ (FWHM) Arecibo beam.  
\label{fig:xray}}
\end{figure}

The high DM is somewhat surprising given the X-ray absorption quoted by
RRK, n$_{\rm H}$=(5.0$\pm0.25)\times 10^{21}$~cm$^{-2}$, where the errors
represent the 90\% confidence region.
The total Galactic HI column density in this direction as estimated from the
FTOOL {\it nh}, which uses the HI map of \citet{dl90}, is $1.2\times 10^{22}$~cm$^{-2}$.  This should be
a good approximation if the source is truly at the far edge of the 
outer spiral arm. Noting that the {\it ASCA} image shows faint, softer emission
in the region (Figure~\ref{fig:xray}), and given the likely possibility of either 
associated thermal X-ray flux from a supernova remnant or a nearby massive star, we fit the {\it ASCA} spectrum of RRK,
adding a thermal component to the absorbed power-law model.
Accounting for $\sim4$\% of the photon flux with a MEKAL thermal plasma model 
of temperature $kT\sim 0.1$~keV in XSPEC \citep{arn96} statistically
improves the fit ({\it F}-test chance probability of 2.5\%). The
best-fit absorption for this three component model is 
n$_{\rm H}$=7.6$\times 10^{21}$~cm$^{-2}$ with a 90\% confidence region
of (4.1 -- 12.3)$\times 10^{21}$~cm$^{-2}$, consistent
with the total Galactic column density. The best-fit photon index
is $\Gamma=1.86$, still consistent with the 1.47 -- 2.01 range in RRK
derived from the simple absorbed power-law model.  Hence the X-ray
absorption does not force us to adopt a smaller distance than is suggested
by the DM.

For a distance $d_{10}=d/10$~kpc, the inferred
isotropic X-ray luminosity $L_X=4.8\times 10^{34} d_{10}^2$ (2 -- 10~keV). 
The X-ray efficiency $\eta_X=L_X/\dot E$ is 0.01$d^2_{10}$.
Compared to 
the total pulsar plus nebula X-ray luminosity of other spin-powered pulsars
this is somewhat high, but within the observed scatter
\citep{pccm02,che00}.

The pulsar's positional coincidence with the error box of the hard spectrum,
low variability {\it EGRET}
$\gamma$-ray source GeV J2020+3658 coupled with the high inferred spin-down 
luminosity strongly suggests this pulsar emits pulsed $\gamma$-rays.
Unfortunately, confirming this by folding archival 
{\it EGRET} data is problematic due to
the likelihood of significant past timing noise and glitches, 
which make the back-extrapolation
of the rotational ephemeris uncertain. RRK noted that the chance 
probability of an X-ray source as bright as AX J2021.1+3651 in the
{\it EGRET} error box was $\sim10$\%, but the nearby Wolf-Rayet star WR141
was equally bright in X-rays and also a potential $\gamma$-ray emitter. 
However, young pulsars remain the only firmly established class of
Galactic {\it EGRET} sources. The known
$\gamma$-ray pulsars cluster at the top of pulsar lists rank-ordered
by spin-down flux $\dot E/d^2$, with $\gamma$-ray efficiencies 
$\eta_\gamma=L_{\gamma}/\dot E$ mostly between 0.001 and 0.03 
(assuming 1 sr beaming) with a tendency
to increase with pulsar age \citep{tbb+99}. The exception
is PSR B1055$-$52, with an apparent $\gamma$-ray efficiency $\eta_\gamma \sim 0.2$
given its nominal DM distance of 1.5 kpc. The inferred $\gamma$-ray efficiency
for PSR J2021+3651 is
$\eta_\gamma =0.15 d_{10}^2$ in the 100~MeV to 10~GeV range. 
If the pulsar is located within the Perseus arm at 
a distance of 5~kpc, then the inferred X-ray and $\gamma$-ray luminosities
would be fairly typical of the other pulsars with Vela-like spin-down 
luminosities. While there is currently no observational evidence for 
a distance this close, increased DM from an
intervening source in this relatively crowded direction would not
be surprising. We note that the DM derived distance for another young pulsar
recently discovered within an $EGRET$ error box, PSR J2229+6114, also
leads to an anomalously high inferred $\gamma$-ray efficiency \citep{hcg+01}. 

\subsection{Upper Limits Toward the Other Sources}

Determining the fraction of radio-quiet versus radio-loud pulsars
is important for our understanding of $\gamma$-ray pulsar emission mechanisms.
The two leading classes of emission models, the outer-gap \citep{rom96a} and 
polar-cap \citep{dh96} models, make very different estimates of the
fraction of $\gamma$-ray pulsars that should be seen at radio energies. 
Out of the 25 brightest sources above 1 GeV not associated with blazars,
$\sim 10$ are now known to either be energetic radio pulsars or contain
such pulsars within their error boxes.
Searching the brightest unidentified X-ray 
sources in five GeV error boxes, we 
detected radio pulsations at the $\sim 0.1$~mJy level (similar to the limiting sensitivity of the 
Parkes observations) from one of these
with Arecibo. This is well below the average flux level expected
for typical radio luminosities of young pulsars \citep{bj99} 
and distances to star forming 
regions statistically associated with $\gamma$-ray sources
\citep{yr97}.
Two of the sources observed with Parkes, AX J1418.7$-$6058 (the Rabbit) and AX J1809.8$-$2333,
have radio and X-ray properties that clearly identify them
as pulsar wind nebulae \citep{rrjg99,brrk02}, and the third, AX J1826.1$-$1300, 
is an extended hard X-ray source that has few other source class
options. Therefore, all three remain viable candidates for $\gamma$-ray
loud, radio-quiet pulsars. Out of this same sample of 25 bright GeV 
sources, the total number of reasonable 
candidate neutron stars within the $\gamma$-ray error boxes
which have now been searched deeply 
for radio-pulsations without success is $\sim 7$.
A current ``best guess" fraction of radio-loud $\gamma$-ray pulsars of 
$\sim 1/2$ falls in between the predictions of the two main competing 
models.

\acknowledgments

We thank Jim Cordes for useful discussions.  We acknowledge support from 
NSERC, CFI, an NSF CAREER Award, and a Sloan Fellowship.  
M.S.E.R. is a Quebec Merit fellow.  S.M.R. is a Tomlinson fellow.  
J.W.T.H. is an NSERC PGS~A fellow, V.M.K. is a Canada Research Chair.  
The Arecibo Observatory is part of the National Astronomy and Ionosphere 
Center, which is operated by Cornell University under a cooperative 
agreement with the National Science Foundation.  The Parkes radio 
telescope is part of the Australia Telescope, which is funded by the 
Commonwealth of Australia for operation as a National Facility 
managed by CSIRO.

\bibliographystyle{apj}
\bibliography{journals1,modrefs,psrrefs,crossrefs,J2020}

\begin{thebibliography}{25}
\expandafter\ifx\csname natexlab\endcsname\relax\def\natexlab#1{#1}\fi

\bibitem[{Arnaud(1996)}]{arn96}
Arnaud, K.~A. 1996, in Astronomical Data Analaysis Software and Systems V, ed.
  G.~Jacoby \& J.~Barnes, Vol. 101 (San Fransisco: ASP), 17

\bibitem[{{Braje} {et~al.}(2002){Braje}, {Romani}, {Roberts}, \&
  {Kawai}}]{brrk02}
{Braje}, T.~M., {Romani}, R.~W., {Roberts}, M.~S.~E., \& {Kawai}, N. 2002,
  \apjl, 565, L91

\bibitem[{Brazier \& Johnston(1999)}]{bj99}
Brazier, K.~T. \& Johnston, S. 1999, MNRAS, 305, 671

\bibitem[Camilo {et~al.}(2001)]{cbm+01}
Camilo, F. et al.  2001, ApJ, 557, L51

\bibitem[{{Camilo} {et~al.}(2002){Camilo}, {Stairs}, {Lorimer}, {Backer},
  {Ransom}, {Klein}, {Wielebinski}, {Kramer}, {McLaughlin}, {Arzoumanian}, \&
  {M{\" u}ller}}]{csl+02}
{Camilo}, F. et al. 2002, \apjl, 571, L41

\bibitem[{{Chevalier}(2000)}]{che00}
{Chevalier}, R.~A. 2000, ApJ, 539, L45

\bibitem[{{D'Amico} {et~al.}(2001){D'Amico}, {Kaspi}, {Manchester}, {Camilo},
  {Lyne}, {Possenti}, {Stairs}, {Kramer}, {Crawford}, {Bell}, {McKay},
  {Gaensler}, \& {Roberts}}]{dkm+01_mal}
{D'Amico}, N. et al. 2001, \apjl, 552, L45

\bibitem[{Daugherty \& Harding(1996)}]{dh96}
Daugherty, J.~K. \& Harding, A.~K. 1996, ApJ, 458, 278

\bibitem[{Dickey \& Lockman(1990)}]{dl90}
Dickey, J.~M. \& Lockman, F.~J. 1990, ARAA, 28, 215

\bibitem[{{Dowd} {et~al.}(2000){Dowd}, {Sisk}, \& {Hagen}}]{dsh00}
{Dowd}, A., {Sisk}, W., \& {Hagen}, J. 2000, in ASP Conf. Ser. 202: IAU Colloq.
  177: Pulsar Astronomy - 2000 and Beyond, 275+

\bibitem[{{Gotthelf} {et~al.}(2000){Gotthelf}, {Ueda}, {Fujimoto}, {Kii}, \&
  {Yamaoka}}]{guf+00}
{Gotthelf}, E.~V., {Ueda}, Y., {Fujimoto}, R., {Kii}, T., \& {Yamaoka}, K.
  2000, \apj, 543, 417

\bibitem[Halpern \& Holt(1992)]{hh92} Halpern, J.~P.~\& 
Holt, S.~S.\ 1992, \nat, 357, 222

\bibitem[{{Halpern} {et~al.}(2001){Halpern}, {Camilo}, {Gotthelf}, {Helfand},
  {Kramer}, {Lyne}, {Leighly}, \& {Eracleous}}]{hcg+01}
{Halpern}, J.~P., {Camilo}, F., {Gotthelf}, E.~V., {Helfand}, D.~J., {Kramer},
  M., {Lyne}, A.~G., {Leighly}, K.~M., \& {Eracleous}, M. 2001, ApJ, 552, L125

\bibitem[Hartman et al.(1999)]{hbb+99} Hartman, R.~C.~et al.\ 
1999, \apjs, 123, 79 

\bibitem[{Helfand(1994)}]{hel94}
Helfand, D.~J. 1994, MNRAS, 267, 490

\bibitem[Lamb \& Macomb(1997)]{lm97} Lamb, R.~C.~\& Macomb, 
D.~J.\ 1997, \apj, 488, 872

\bibitem[{Manchester {et~al.}(2001)Manchester, Lyne, Camilo, Bell, Kaspi,
  D'Amico, McKay, Crawford, Stairs, Possenti, Morris, \& Sheppard}]{mlc+01}
Manchester, R.~N. et al. 2001, \mnras, 328, 17

\bibitem[{Manchester \& Taylor(1977)}]{mt77}
Manchester, R.~N. \& Taylor, J.~H. 1977, Pulsars (San Francisco: Freeman)

\bibitem[{Nice \& Sayer(1997)}]{ns97}
Nice, D.~J. \& Sayer, R.~W. 1997, ApJ, 476, 261

\bibitem[{{Possenti} {et~al.}(2002){Possenti}, {Cerutti}, {Colpi}, \&
  {Mereghetti}}]{pccm02}
{Possenti}, A., {Cerutti}, R., {Colpi}, M., \& {Mereghetti}, S. 2002, \aap,
  387, 993

\bibitem[{{Ransom}(2001)}]{ran_thesis}
{Ransom}, S.~M. 2001, Ph.D.~Thesis

\bibitem[{Roberts {et~al.}(1999)Roberts, Romani, Johnston, \& Green}]{rrjg99}
Roberts, M. S.~E., Romani, R.~W., Johnston, S., \& Green, A.~J. 1999, ApJ, 515,
  712

\bibitem[Roberts, Romani, \& Kawai(2001)]{rrk01}
Roberts, M. S.~E., Romani, R.~W., \& Kawai, N. 2001, ApJS, 133, 451

\bibitem[{Romani(1996)}]{rom96a}
Romani, R.~W. 1996, ApJ, 470, 469

\bibitem[{Taylor \& Cordes(1993)}]{tc93}
Taylor, J.~H. \& Cordes, J.~M. 1993, ApJ, 411, 674

\bibitem[{Thompson {et~al.}(1999)Thompson, Bailes, Bertsch, Cordes, D'Amico,
  Esposito, Finley, Hartman, Hermsen, Kanbach, Kaspi, Kniffen, Kuiper, Lin,
  Manchester, Matz, Mayer-Hasselwander, Michelson, Nolan, Ogelman, Pohl,
  Ramanamurthy, Sreekumar, Reimer, Taylor, \& Ulmer}]{tbb+99}
Thompson, D.~J. et al. 1999, ApJ, 516, 297

\bibitem[Thompson(2001)]{tho01} Thompson, D.~J.\ 2001, ASSL 
Vol.~267: The Nature of Unidentified Galactic High-Energy Gamma-Ray 
Sources, 3

\bibitem[{{Voges} {et~al.}(2000){Voges}, {Aschenbach}, {Boller}, {Brauninger},
  {Briel}, {Burkert}, {Dennerl}, {Englhauser}, {Gruber}, {Haberl}, {Hartner},
  {Hasinger}, {Pfeffermann}, {Pietsch}, {Predehl}, {Schmitt}, {Trumper}, \&
  {Zimmermann}}]{vab+00}
{Voges}, W., et al. 2000, VizieR Online
  Data Catalog, 9029, 0+

\bibitem[{Yadigaroglu \& Romani(1997)}]{yr97}
Yadigaroglu, I.-A. \& Romani, R.~W. 1997, ApJ, 476, 347

\end{thebibliography}

\begin{deluxetable}{c c c c c c c c c c c}
\rotate
\tablewidth{0pt}
\tabletypesize{\scriptsize}
\tablecaption{Observational Parameters \label{tab:observations}}
\tablehead{\colhead{Source} & \colhead{RA (J2000)} & \colhead{DEC (J2000)} & \colhead{Epoch} & \colhead{Telescope} & 
\colhead{$\nu_c$\tablenotemark{a}} & \colhead{$\Delta\nu$\tablenotemark{b}} & \colhead{N$_{\rm ch}$\tablenotemark{c}} & 
\colhead{t$_{\rm sam}$\tablenotemark{d}} & \colhead{T$_{\rm int}$\tablenotemark{e}} & \colhead{$S_{\rm min}$\tablenotemark{f}} \\
\colhead{Name} & \colhead{hh:mm:ss.s} & \colhead{dd:mm:ss} &\colhead{(MJD)} & & \colhead{(MHz)} & \colhead{(MHz)} &  &
\colhead{($\mu$s)} & \colhead{(s)} & \colhead{(mJy)}} 

\startdata
AX J1418.7$-$6058 / GeV J1417$-$6100 & 14:18:41.5 & $-$60:58:11 & 51951.63 & Parkes  & 1390 & 256 & 512 & 250 & 16900 & 0.08\\
                  &  &  & 51953.67 & Parkes & 1373 & 288 &  96 & 250 & 16900 & 0.08\\

AX J1809.8$-$2332 / GeV J1809$-$2327  & 18:09:50.2 & $-$23:32:23 & 51951.83 & Parkes  & 1390 & 256 & 512 & 250 & 16900 & 0.08\\
                  &  &  & 51954.77 & Parkes & 1373 & 288 &  96 & 250 & 16900 & 0.08\\

AX J1826.1$-$1300 / GeV J1825$-$1310  & 18:26:04.9 & $-$12:59:48 & 51952.83 & Parkes  & 1390 & 256 & 512 & 250 & 13234 & 0.09 \\
                  &  &  & 51955.83 & Parkes & 1373 & 288 &  96 & 250 & 14832 & 0.09 \\

AX J1907.4+0549 / GeV J1907+0557 & 19:07:21.3 & +05:49:14 & 52305.58 & Arecibo & 1425 & 100 & 512 & 200 & 6480 & 0.02\\

AX J2021.1+3651 / GeV J2020+3658  & 20:21:07.8 & +36:51:19 & 52304.67 & Arecibo & 1425 & 100 & 512 & 200 & 1627 & 0.04 \\
                  &  &  & 52305.66 & Arecibo & 1425 & 100 & 256 & 200 & 3000 & 0.03 \\

\enddata
\tablenotetext{a}{Central observing frequency.}
\tablenotetext{b}{Total bandwidth of the observation.}
\tablenotetext{c}{Number of frequency channels.}
\tablenotetext{d}{Sampling time.}
\tablenotetext{e}{Length of the observation.}
\tablenotetext{f}{Flux density sensitivity limit.}

\end{deluxetable}

\begin{deluxetable}{l c}
\tabletypesize{\scriptsize}
\tablewidth{0pt}
\tablecaption{Measured and Derived Parameters for PSR~J2021+3651 \label{tab:pulsar}}
\tablehead{\colhead{Parameter} & \colhead{Value}}

\startdata

Right ascension $\alpha$ (J2000) & 20$^{\rm{h}}$ 21$^{\rm{m}}$ 
04$^{\rm{s}}$.5\tablenotemark{a} 
\\
Declination $\delta$ (J2000) & +36$^\circ$ 51$'$ 27$''$.0\tablenotemark{a}\\
Galactic latitude  $l$          & 75.$^{\circ}$23 \\
Galactic longitude $b$          & +0.$^{\circ}$11 \\
Pulse period $P$ (s)         & 0.10372222480(17) \\
Period derivative  $\dot P$     & $ 9.563(48) \times 10^{-14}$  \\  
Pulse frequency $\nu$ (s$^{-1}$) & 9.641135271(16) \\         
Frequency derivative  $\dot{\nu}$ (s$^{-2}$) & $-8.889(45) \times 10^{-12}$ \\
Epoch (MJD)        & 52407.389 \\
Dispersion Measure DM (pc cm$^{-3}$) & 371(3) \\
Pulse width at 50\% of peak w$_{50}$ (ms)       & 9.9  \\
Pulse width at 10\% of peak w$_{10}$ (ms)       & 18 \\
Flux density at 1425 MHz (mJy)     & $\sim 0.1$ \\                                       
Spin-down luminosity $\dot{E}$\tablenotemark{b} (erg s$^{-1}$) & $3.4 \times 10^{36}$ \\
Surface dipole magnetic field $B$\tablenotemark{c} (G)  & $3.2 \times 10^{12}$ \\
Characteristic Age $\tau_{c} \equiv \frac{1}{2} P / \dot P$ (kyr)     &  17 \\

\enddata

\tablecomments{ Figures in parentheses represent uncertainty in the least-significant digits quoted, equal to 3 times errors given by {\tt TEMPO}.}

\tablenotetext{a}{ Coordinates of the $ROSAT$ X-ray source 
1RXS J202104.5+365127 with estimated positional error of $24^{''}$.}

\tablenotetext{b}{ $\dot E = 4 {\pi}^2 I \dot P / P^3$ with $I = 10^{45}$ g cm$^2$. }

\tablenotetext{c}{ Assuming standard magnetic dipole spindown:
$B = 3.2 \times 10^{19} (P \dot P)^{1/2}$ Gauss \citep{mt77}.}


\end{deluxetable}

\clearpage

\end{document}